\documentclass[RNAAS]{aastex62}
\usepackage{amssymb}

\def\loggf{\log{gf}}

\begin{document}

\title{Spectral lines in the CaII triplet region\\for RAVE DR6 chemical 
abundance pipeline}

\correspondingauthor{Guillaume Guiglion}
\email{gguiglion@aip.de }

\author[0000-0002-1317-2798]{Guillaume Guiglion}
\affiliation{Leibniz-Institut f\"ur Astrophysik Potsdam (AIP), An der Sternwarte 16, 
14482 Potsdam, Germany}

\author[0000-0003-1269-7282]{Cristina Chiappini}
\affiliation{Leibniz-Institut f\"ur Astrophysik Potsdam (AIP), An der Sternwarte 16, 
14482 Potsdam, Germany}

\author[0000-0003-0974-4148]{Marica Valentini}
\affiliation{Leibniz-Institut f\"ur Astrophysik Potsdam (AIP), An der Sternwarte 16, 
14482 Potsdam, Germany}

\author[0000-0001-6516-7459]{Matthias Steinmetz}
\affiliation{Leibniz-Institut f\"ur Astrophysik Potsdam (AIP), An der Sternwarte 16, 
14482 Potsdam, Germany}


\section{} 

The purpose of this research note is to present a reliable list of clean spectral lines in the 
Calcium Triplet region at medium resolution $R\sim7\,500$, for chemical abundance analysis. 
Such line lists present a general interest for on-going large spectroscopic survey, such as 
Gaia-RVS \citep{bailer_jones_2013}, 4MOST \citep{de_jong_2014} and WEAVE \citep{dalton_2014}. 
The framework of the present study is the RAdial Velocity Experiment (RAVE), targeting roughly 
half-million of stars \citep{steinmetz_2006} thanks to the multi-object spectrograph 6dF 
(6 degree field) of the 1.2 m UK Schmidt Telescope of the Australian Astronomical 
Observatory (AAO). The wavelength spectral coverage is centered on the Infrared Calcium 
Triplet $\lambda\in[8\,410-8\,795]$\AA, with an effective medium-resolution resolution 
of $R\sim7\,500$. 

The RAVE spectra have relatively small wavelength coverage ($\sim400$\AA), but possess 
sufficient spectral signatures in order to derive chemical abundances of three 
$\alpha-$elements Magnesium, Silicon and Titanium, two iron-peak elements Iron and 
Nickel, and finally Aluminum. For the upcoming sixth Data Release of RAVE (Steinmetz 
et al. in preparation), a new dedicated automatic chemical abundances pipeline was 
adapted to the spectral range and resolution of RAVE spectra: GAUGUIN \citep{guiglion_2016}. 
This method already showed its efficiency in many applications: $\alpha-$ and iron-peak 
elements abundances at medium resolution for the Gaia-ESO Survey (ex: \citealt{rojas_2016}), 
and high-resolution for the Gaia benchmark stars \citep{jofre_2015}, lithium \citep{guiglion_2016} 
and \textit{s-/r-}process abundances at high resolution with ESO Achives spectra 
\citep{guiglion_2018}.

In this research note we focus on spectral line selection in order to retrieve the 
best chemical abundances. We proceeded to a careful examination of the spectral lines 
available on the RAVE wavelength domain. We took advantage of the line list and synthetic spectra 
grid intensively used in RAVE for the atmospheric parameter determination 
\citep{kordopatis_2011, kordopatis_2013, kunder_2017}. We recall that the grid was 
computed using the MARCS atmosphere models \citep{gustafsson_2008} and the LTE 
TURBOSPECTRUM code \citep{plez_2012}.This grid, at RAVE spectral resolution, covers 
the domain $\lambda\in[8410-8795]$\AA~with a step $\Delta \lambda = 0.4\,$\AA.
For a given line, we validated its selection 
comparing by eye the observed spectra of the Sun and Arcturus (\citet{hinkle_2003}, both 
at $R\sim100\,000$ and $R\sim7\,500$) with respect their respective synthetic spectra 
(interpolated in the quoted above grid). Spectral lines which were too blended or too weak were rejected. 
In total, 39 spectral lines were retained, see \tablename~\ref{table_line}. 

We note that the abundances derived with such a line list on RAVE spectra provide very 
satisfactory results (\citet{Guiglion_2018_rave}, Steinmetz et al. in preparation). 
We notice that such a line list is especially adapted for RAVE resolution and 
wavelength coverage. It can be a starting point for spectral abundance analysis in 
other on-going large spectroscopic survey (ex: Gaia-RVS, 4MOST, WEAVE) but will require 
more specific look regarding the resolution, and complementary lines in case of higher 
resolution or larger wavelength coverage.

\begin{table*}
\centering
\begin{tabular}[c]{c c c c}
El. & line (\AA) & $\chi_e$ & $\loggf$ \\
\hline
\hline
MgI & $8\,712.682$ & 5.932 & -1.18 \\
MgI & $8\,717.815$ & 5.933 & -0.87 \\
MgI & $8\,736.016$ & 5.946 & -0.26 \\
AlI & $8\,772.865$ & 4.022 & -0.39 \\
AlI & $8\,773.897$ & 4.022 & -0.20 \\
SiI & $8\,443.970$ & 5.871 & -1.38 \\
SiI & $8\,555.903$ & 5.616 & -2.39 \\
SiI & $8\,556.777$ & 5.871 & -0.35 \\
SiI & $8\,556.805$ & 6.191 & -0.55 \\
SiI & $8\,648.465$ & 6.206 & +0.01 \\
SiI & $8\,742.446$ & 5.871 & -0.54 \\
SiI & $8\,752.007$ & 5.871 & -0.40 \\
TiI & $8\,426.506$ & 0.826 & -1.37 \\
\hline
\end{tabular}
\begin{tabular}[c]{c c c c}
El. & line (\AA) & $\chi_e$ & $\loggf$ \\
\hline
\hline
TiI & $8\,434.957$ & 0.848 & -0.99 \\
TiI & $8\,435.652$ & 0.836 & -1.01 \\
TiI & $8\,518.028$ & 2.134 & -1.06 \\
TiI & $8\,518.352$ & 1.879 & -0.89 \\
TiI & $8\,682.980$ & 1.053 & -2.02 \\
TiI & $8\,692.331$ & 1.046 & -2.23 \\
TiI & $8\,734.712$ & 1.053 & -2.35 \\
TiI & $8\,766.676$ & 1.067 & -2.39 \\
FeI & $8\,514.794$ & 5.621 & -2.13 \\
FeI & $8\,515.108$ & 3.018 & -2.13 \\
FeI & $8\,526.669$ & 4.913 & -0.71 \\
FeI & $8\,582.257$ & 2.990 & -2.36 \\
FeI & $8\,592.951$ & 4.956 & -0.91 \\
\hline
\end{tabular}
\begin{tabular}[c]{c c c c}
El. & line (\AA) & $\chi_e$ & $\loggf$ \\
\hline
\hline
FeI & $8\,611.803$ & 2.845 & -2.06 \\
FeI & $8\,621.601$ & 2.949 & -2.47 \\
FeI & $8\,688.624$ & 2.176 & -1.33 \\
FeI & $8\,698.706$ & 2.990 & -3.32 \\
FeI & $8\,699.454$ & 4.955 & -0.54 \\
FeI & $8\,713.187$ & 2.949 & -3.08 \\
FeI & $8\,713.208$ & 4.988 & -1.04 \\
FeI & $8\,757.187$ & 2.845 & -2.09 \\
FeI & $8\,763.966$ & 4.652 & -0.33 \\
NiI & $8\,579.978$ & 5.280 & -0.94 \\
NiI & $8\,636.995$ & 3.847 & -1.94 \\
NiI & $8\,702.489$ & 2.740 & -3.19 \\
NiI & $8\,770.672$ & 2.740 & -2.79 \\
\hline
\end{tabular}
\caption{\label{table_line}Element, wavelength, excitation potential 
and $\loggf$ of the spectral lines used in RAVE DR6 for the chemical 
abundance pipeline GAUGUIN.}
\end{table*}

\acknowledgments

\end{document}